\documentclass[12pt]{article}
\usepackage{amsmath,amssymb}

\begin{document}
\title{\Large\bf On Credit-based Incentive Mechanisms of Voluntary User Comment Reviewing in Social Networks}
\author{Shiyu Ji}
\date{}
\maketitle

\begin{abstract}
With the recent advance of micro-blogs and social networks, people can view and post comments on the websites in a very convenient way.
However, it is also a big concern that the malicious users keep polluting the cyber environment by scamming, spamming or repeatedly advertising.
So far the most common way to detect and report malicious comments is based on voluntary reviewing from honest users. 
To encourage contribution, very often some non-monetary credits will be given to an honest user who validly reports a malicious comment. 
In this note we argue that such credit-based incentive mechanisms should fail in most cases:
if reporting a malicious comment receives diminishing revenue, then in the long term no rational honest user will participate in comment reviewing.
\end{abstract}

\section{Introduction}
It is out of question today is in the era people can easily share their own opinions on social websites \cite{E07,ECL07}. Unfortunately this great advancement also attracts malicious users who keep posting false or meaningless information online for their own benefits. For example, sybil attacks recently develop very fast and receive a lot of attention from the researchers \cite{D02,N04}. So far the most common countermeasure against malicious comment pollution is to rely on the power from the honest users, who are usually the majority. Honest users will not only obey the online community rules, but also faithfully report any violating comments they have read to the administrators. Assuming most users are honest, a voluntary comment reviewing scheme can be built. This is almost a basic feature of every comment-sharing or social website today.

However, it is clear that a rational honest user will not actively participate in a purely voluntary reviewing system: finding violating comments and writing reports require human effort, and very often the content of malicious comments shall make the readers unpleasant. Doing comment reviewing always incurs cost. If the cost is too much to tolerate, the honest users will not read the comments any more: they can leave away or block related comments from their devices. If there is no incentive to the honest users, most rational people cannot contribute reports to the administrators in the long term. Thus an incentive mechanism is needed to encourage involvement, and in fact some websites have tried this idea.

Social websites like Weibo \cite{WB}, Tencent Qzone \cite{QZ} and micro-blog \cite{TMB}, Bilibili \cite{BL} use a reputation-based system to encourage user comment reviewing.
When the users validly submit reports on violating comments and get processed by administrators, they shall receive some credits as a proof of contributed effort and website-wide reputation. These incentive mechanisms share the features as follows:
\begin{itemize}
\item The given credits cannot be refunded or join the physical market. The user credits can only be consumed in the website only, e.g., unlocking new features, paying for violating rules, etc.
\item There is no competition between the users who have the credits, i.e., usually the information about their credits is confidential.
\item There is a maximum amount of credits a user can get. After achieving the maximum, the user can still report, but will not gain anything from reporting.
\end{itemize}
However, as we will show, these credit-based incentive mechanisms can barely work. In fact, Tencent has closed its incentive system for years. Bilibili is still using credits as incentives, but the achievement is far less than expectation \cite{ZH16}. What is the reason that such a system fails?

It turns out the major problem is that the revenue for each report diminishes as the user submits more and more reports. Since the credits are only for reputation and website use, they cannot be refunded or turn into any monetary form. As more and more users join the website, the value of credits will decrease since 1) there are more and more credits distributed to the users, and 2) a new user is often given some initial credits. These will cause inflation-like phenomenon on the credits. However, to submit an honest report always requires a certain amount of human effort. When the reporting cost is larger than the diminishing revenue, any rational honest users shall refuse to continue contributing since their utilities are in deficit. We will explain more details in the following section.

\section{The Game When Reviewing a User Comment}
Upon receiving a comment from another user, an honest user has three options to choose: 1) read the comment only, 2) read the comment and report to an administrator if it violates community rules, and 3) simply discard the comment. Clearly if the user discards the comment, she will not gain or lose anything, and hence receives zero utility $U_3 = 0$. We need to analyze the rest two cases.
\subsection{Notations and Preliminaries}
\begin{itemize}
\item $c_r$ \--- the human effort/cost to read a comment
\item $p$ \--- the probability that a comment is malicious
\item $c_p$ \--- the expected mental cost of a malicious comment
\item $c_w$ \--- the cost to write and submit an honest report
\item $r$ \--- the number of credits to give when an honest report is processed by administrator
\item $u$ \--- the unit revenue of one credit
\item $R_i$ \--- (physical or mental) revenue for choice $i$
\item $C_i$ \--- (physical or mental) cost for choice $i$
\item $U_i$ \--- (physical or mental) utility for choice $i$
\end{itemize}
In game theory, the utility $U_i$ upon strategy $i$ is defined as $U_i = R_i - C_i$.

\subsection{Computing Utilities}
\begin{itemize}
\item If the user chooses to only read the comment, then with probability $p$ she will meet a malicious one and thus suffer a mental cost $c_p$, and with probability $1-p$ she will meet a good one. Thus the expected cost $C_1$ of this strategy shall be
$$C_1 = c_r + p\cdot c_p.$$
Note that there is no revenue to the user since she did not report or involve in reviewing. Hence for this choice, the utility is
$$U_1 = -C_1 = -c_r - p\cdot c_p.$$
\item If the user chooses to read and report, then with probability $p$ she will meet a malicious one, pay a mental cost, and thus write a report, and with probability $1-p$ she will see a good comment and nothing will happen. Thus the expected cost $C_2$ of this choice shall be
$$C_2 = c_r + p\cdot(c_p + c_w).$$
For the revenue, with probability $p$ the user will receive $r$ credits for the submitted report, while with probability $1-p$ there is no credit. Thus the expected revenue shall be 
$$R_2 = p\cdot r\cdot u.$$
Hence the utility of this choice is
$$U_2 = R_2 - C_2 = -c_r + p\cdot(r\cdot u -c_p-c_w).$$
\end{itemize}

\subsection{Discussion on Strategy Choice}
Clearly there is a dominating strategy: $U_1 < 0 = U_3$, i.e., a rational honest user shall not view a malicious comment without giving any feedback or reporting. Passively receiving negative input will only cause unhappiness or waste of time. It is much better to simply discard the comments which are potentially malicious. Hence we only need to discuss the choice between 2) and 3).

Any rational user chooses strategies based on utilities, i.e., she shall choose 2) if $U_2 > U_3$, and otherwise shall choose 3). If $U_2 > U_3$, then
$$-c_r + p\cdot(r\cdot u -c_p-c_w) > 0,$$
$$p\cdot r\cdot u > c_r+p\cdot(c_p+c_w).$$
Thus the condition that a rational user will choose to read and report is:
$$u > \frac{c_r}{p\cdot r} + \frac{c_p+c_w}{r}.$$
Note that if the malicious probability $p$ is very small, then $u$ needs to be very large to maintain the incentive mechanism. In this case the online social community has a good environment ($p$ is small), and thus there is no point to sustain a reviewing scheme until a group of malicious users appear.

Given a non-negligible $p$, we expect the number of credits $r$ to be as large as possible to make $u$ satisfy the condition above. However, since reputation-based credits has no proof of work \cite{PoW}, they cannot be redeemed as money like Bitcoin \cite{BCoin}. With more credits to return, the unit revenue $u$ will change proportionally to $1/N$ where $N$ is the total amount of distributed credits. Even though the condition can be satisfied temporarily, after some time it will not hold any more since $u$ keeps decreasing as the distributed credits are accumulated. This is very similar to inflation in economics. Also we note that a newly registered user is usually given a certain amount of credits as the initial incentive. This policy aggravates the credit inflation problem.

If there is a maximum amount of credits a user can get, then after reaching the maximum, the honest user will not gain anything. Then the unit revenue $u$ becomes zero. Thus the condition above can never be satisfied. Any incentive mechanism with a credit bound asks honest users to leave after they reach the full rank.

In general, the revenue function $R(n)$, where $n$ is the number of valid reports submitted, is submodular \cite{RS09}. The marginal revenue $\Delta R(n) := R(n+1)-R(n)$ decreases as $n$ grows, e.g., $R(n) \propto \log{n}$. Thus the unit revenue $u(n)$ is
$$u(n) = \frac{R(n)}{n} \propto \frac{\log{n}}{n}.$$
Suppose $R(n) = \alpha\log{n}+\beta$ for some $\alpha\in\mathbb{R}^+$ and $\beta\in\mathbb{R}$. Then the limit of unit revenue is zero.
$$\lim_{n\to +\infty}u(n) = \lim_{n\to +\infty}\frac{\alpha\log{n}+\beta}{n} = 0.$$
This implies in the cases of submodular revenue (very common in the real world), there is no way to keep any rational honest user contributing valid reports in the long term only by using credit-based incentives.

Hence in the long term, the rational honest users shall simply discard all the user comments from the website. They will just read the news, watch the video and nothing more. They make such a choice just because they do not want deficit. While the honest users quit, the malicious users are still active when making comments. This will further deteriorate the commentary environment.

\section{Remarks on Irrational Users}
An important assumption in this note is that the honest users are \emph{rational}. However in the real world, some honest users will do something irrational, e.g., getting into a comment-war with a malicious user, wasting time on useless comments for nothing, etc. For the malicious users, it is more difficult to predict their choices. Usually it is very difficult to model irrational entities. More subtle models and theories may be left as future work.

\section{Conclusion}
This note explains 1) why many senior and honest users eventually quit the online commentary community, becoming increasingly closed against comments from others, and 2) why the reputation or credit-based incentive mechanisms fail to motivate the users to review comments in the long term. As long as the credits cannot be refunded into monetary form, it is not a good idea to base such an incentive mechanism, which motivates reporting user comments, solely on credits.

Shiyu Ji: Department of Computer Science, UCSB.

Email: shiyu@cs.ucsb.edu
\end{document}